\documentclass[conference]{IEEEtran}

\ifCLASSINFOpdf
\else
\fi

\usepackage{amsmath}
\usepackage{amssymb}
\usepackage{array}
\usepackage{graphicx}
\usepackage{color}
\usepackage{cite}
\usepackage{adjustbox}

\begin{document}

\title{Multiscale Fractal Analysis on EEG Signals \\ for Music-Induced Emotion Recognition}



\author{
\IEEEauthorblockN{
Kleanthis Avramidis,
Athanasia Zlatintsi,
Christos Garoufis and
Petros Maragos
}
\vspace{0.15cm}
\IEEEauthorblockA{\textit{School of ECE, National Technical University of Athens}, 15773 Athens, Greece \\
kle.avramidis@gmail.com, cgaroufis@mail.ntua.gr, \{nzlat, maragos\}@cs.ntua.gr}
}

\maketitle

\begin{abstract}

Emotion Recognition from EEG signals has long been researched as it can assist numerous medical and rehabilitative applications. However, their complex and noisy structure has proven to be a serious barrier for traditional modeling methods. In this paper, we employ multifractal analysis to examine the behavior of EEG signals in terms of presence of fluctuations and the degree of fragmentation along their major frequency bands, for the task of emotion recognition. In order to extract emotion-related features we utilize two novel algorithms for EEG analysis, based on Multiscale Fractal Dimension and Multifractal Detrended Fluctuation Analysis. The proposed feature extraction methods perform efficiently, surpassing some widely used baseline features on the competitive DEAP dataset, indicating that multifractal analysis could serve as basis for the development of robust models for affective state recognition.

\end{abstract}


\vspace{-0.1cm}
\section{Introduction}

Machine Learning has made overwhelming progress in modeling rational intelligence and the way humans perceive and act upon their environment. In tasks such as Object Recognition \cite{obj_rec} and Sequence Prediction \cite{NIPS2018_8004}, supervised machine learning systems have accomplished to surpass, in many cases, the human brain capabilities. However, there are still many challenges in approaching emotion-driven intelligence, although it constitutes a fundamental aspect of human's perception and decision-making processes. The reason for this is that emotions are highly subjective, and thus really difficult to be labeled when expressed. A large number of studies attempt to include speech \cite{speech1, speech2}, text \cite{AbdulMageed2017EmoNetFE}, as well as facial expressions \cite{Fernandez_2019_CVPR_Workshops} in building emotion recognition systems. Nevertheless, there is a growing interest in emotion tagging through physiological signals \cite{phys}, since those are induced without our active interference and thus depict more clearly the actual affective state. Such methods have been popular in designing Human-Computer or Brain-Computer Interfaces (BCI) that aid humans and adapt to personalized preferences. Further, physiological signals have been used for medical purposes, among others for the detection of epilepsy \cite{epilepsy} and depression \cite{depression}. 

Among a variety of physiological signals, special consideration is given to brain data. The electroencephalogram (EEG) is the most widely researched signal of its kind and has been highly effective in detecting affective states. A variety of time \cite{timefeats}, frequency \cite{Wang2011EEGBasedER} and joint \cite{wavelets} domain features have been extracted from EEG. Particular attention has been given to channel connectivity features, such as mutual information \cite{mi} and differential asymmetry \cite{dasm}, reporting the highest scores in literature. Nowadays though, various types of deep neural networks have exceeded the performance of traditional feature-oriented methods \cite{bihemispheric, 3dcnn}.

However, processing EEG signals and extracting useful features remain core challenges, since EEG, like most biological signals, is chaotic, nonlinear and incorporates a large amount of noise, both from the recording equipment and interfering physiological processes \cite{KUMAR20122525}. Because of the nature of such signals \cite{doi:https://doi.org/10.1002/9780470511923.ch7} several nonlinear fractal methods have been proposed, one of them being the Higuchi Fractal Dimension \cite{HIGUCHI1988277}, which has been used extensively in emotion recognition as an analysis tool \cite{fractal1, fractal2}. Due to their complexity though, such signals do not always share the same structure over every time scale, hence the fractal characteristics may vary and change dynamically or accordingly to the examined scale. For this reason, in this paper we propose the Multiscale Fractal Dimension \cite{MARAGOS1994199} and Multifractal Detrended Fluctuation Analysis \cite{KANTELHARDT200287} to examine the EEG signals and determine emotional information buried in their fragmented structure. In order to demonstrate the efficiency of the proposed multifractal features, we modify the BCI workflow \cite{eeg_general} to include our Feature Engineering algorithms (Fig.~1), obtaining competitive results that surpass some widely used baseline methods.

The rest of the paper is organized as follows: Sec.~2 provides a detailed description of the multifractal algorithms used in this study. In Sec.~3, we analyze the structure of EEG signals in terms of stationarity and fragmentation, which are important factors in fractal algorithms. Section~4 describes the feature extraction methodology, whereas in Sec.~5 we describe the the context of our experiments and discuss their outcomes. Finally, in Sec.~6 we conclude our analysis and propose further directions for future work on the field.

\begin{figure}[]
    \centering
    \includegraphics[scale=0.41]{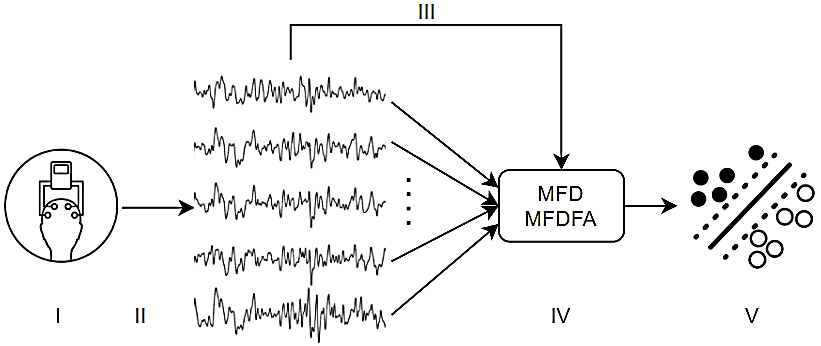}
    \caption{Experiment Pipeline: I) EEG acquisition \cite{KOELSTRA} II) EEG Preprocessing \cite{KOELSTRA} III) Stationarity and Fragmentation Analysis IV) Multifractal Feature Extraction V) Binary Classification of concatenated Features}
    \vspace{-0.5cm}
\end{figure}

\section{Multifractal Methods}
\label{sec:format}

In signal analysis and geometry, the Fractal Dimension $D$ is an index of complexity and fragmentation, comparing how the details of a signal's pattern change when measured at different scales. The fractal dimension of 1D signals may vary between 1 and 2, and the larger the $D$ is, the larger the amount of fragmentation of the signal.
Alternatively, one could consider 
the Hurst Exponent $H$ of a signal to analyse the global properties of its fluctuations. Given a self-similar signal, e.g. fractional brownian motion, the fractal dimension is derived from the Hurst exponent through $H = 2-D$.

\subsection{Multiscale Fractal Dimension}

Maragos \cite{MARAGOS1994199} developed an efficient algorithm to determine the Multiscale Fractal Dimension (MFD) of a signal by measuring the multiscale length of a curve with disks of varying radius via morphological coverings. The cover is created using 2D morphological set dilations of the signal graph $F$ by multiscale versions $sB = \{sb : b \in B\}$ of a unit-scale convex symmetric set $B$, $s\geq0$ the scale:
$$ F \oplus sB = \{z+sb \in \mathbb{R}^2 : z \in F, b \in B\}. $$
Then, the cover area $A_B(s) = \mathrm{area}(F \oplus sB)$ is computed and the fractal dimension $D$ is yielded by:
%
$$ D = \lim_{s \rightarrow 0}{\frac{\log[A_B(s)/s^2]}{\log[1/s]}}. $$
It has been shown \cite{MARAGOS1994199} that the above limit will not change if we approximate $A_B(s)$ with 1D nonlinear convolutions instead of 2D set operations,  which enables its efficient calculation. In practice, $D$ can be estimated by a least-squares line to find the slope of $\log[A_B(s)]$ versus $\log(s)$ since
%
$$ \log[A_B(s)] = (2-D)\log(s) + \mathrm{constant},$$
assuming the power law $A_B(s) \approx s^{2-D}$ as $s\rightarrow 0$. We therefore compute the slope of the data over a small scale window of $w$ scales that move along the scale axis $s$ $\{s, s+1, ..., s+w\}$, creating a profile of local MFDs $D(s,t)$ at each time $t$ (fractogram). The local slope is now an estimate of $2-D$ and from this the fractal dimension $D$ can be easily derived.

\subsection{Detrended Fluctuation Analysis}

Detrended Fluctuation Analysis (DFA) \cite{article} estimates the Hurst exponent $H$ in time series data $x[n]$ of length $N$ by utilizing its cumulative sum $y[n] = \sum_{n=1}^N{(x[n]-\mu_x)}$. This profile is divided into $N_s$ non-overlapping windows $y[k,n], k=1,...,N_s$ of length $s$ and for every window the local trend $r[k,n]$ is obtained through linear regression. We denote $y_d[k,n] = y[k,n] - r[k,n]$ the detrended version of the k-th profile segment. Then, the RMS value of each detrended segment is computed and averaged across the segments:
$$F(s) = \sqrt{\frac{1}{N_s}\sum_{k=1}^{N_s}{F_k^2(s)}} , \quad F_k(s) = \sqrt{\frac{1}{s}\sum_{n=1}^{s}{y_d[k,n]^2}}. $$
The result of the above operations is a vector of $s$ values, one for each chosen scale. The relationship between $F(s)$ and $s$ is described by the power law $ F(s) \propto s^H $, which determines $H$. Multifractal DFA (MFDFA) \cite{KANTELHARDT200287} is essentially a generalization of DFA, where the computation of $F(s)$ includes $q$ moments:
\vspace{-0.2cm}
$$ F_q(s) = \sqrt[q]{\frac{1}{N_s}\sum_{k=1}^{N_s}{F_k^q(s)}}.$$
As a result, a separate line is computed for every value of the factor $q$, with $q=2$ being the reduction to classical DFA. MFDFA could prove especially useful in cases where the scaling exponents and complexities are dependent on the scale, or change dynamically, in the context of time series.

\section{EEG Signals}

\subsection{Dataset}

We work on the competitive, but widely used DEAP Dataset \cite{KOELSTRA}, including data from 32 subjects in their already preprocessed form. Each subject is exposed to forty 60-seconds long music videos as stimuli, while having their EEG recorded, along with other physiological signals. After watching each video, the subject was instructed to rate their induced emotion in 5 dimensions: valence (pleasantness), arousal (excitation), dominance, liking and familiarity to the stimulus. We solely experiment with valence and arousal, as they form a complete emotion space \cite{KOELSTRA}. The rating ranges from 1 (weakest) to 9 (strongest). The EEG signals were recorded at a sampling rate of 512 Hz and downsampled to 128 Hz. The 10-20 placement system was followed, using 32 electrodes.
 
\subsection{Stationarity \& Hurst Exponent Estimation}
\label{sssec:subsubhead}

Physiological signals like the EEG are widely researched as noisy and non-stationary signals and commonly demand heavy pre-processing. The observed structure is partly due to external stimuli or other physiological operations and mainly indicates the complexity and the states of neural assemblies during brain functioning \cite{KUMAR20122525}. In our experiments, it is crucial to determine the stationarity of the signals in order to correctly interpret their fractal properties. We apply the Augmented Dickey-Fuller (ADF) Test to a randomly sampled set of EEG signals and, to our surprise, we derive evidence of strict stationarity. Specifically, the examined signals appeared non-stationary only at very low scales, up to windows of 100 samples or 0.8 seconds. The same holds when we test the signal profiles, i.e. the cumulative sums. However, a few signals exhibit non-stationarities at their major frequency bands. In order to find the source of this stationarity, we reproduced the preprocessing applied in \cite{KOELSTRA} to a sample raw waveform. This included downsampling to 128 Hz, eye-artefact removal, filtering at 4--45Hz and averaging to the common reference channel. After this procedure, it was found that the cause of stationarity was the performed bandpass filtering.

Although fractal methods are used for analyzing time series that appear to have long-memory correlations and non-stationary dynamics, they are not restricted to such. Scale-free stationary processes can be viewed as \textit{fractional Gaussian noise} (fGn), while their increments typically construct non-stationary processes in the form of \textit{fractional Brownian motion} (fBm), of the same Hurst exponent. Thus, the exponent estimation is crucial in characterizing EEG signals for multifractal analysis \cite{10.3389/fphys.2012.00141} and can be determined by monofractal DFA. If the estimated exponent is less than $H=1$, then it characterizes a stationary process, which can be modeled as fGn with that exponent. Otherwise, it is assumed to be produced by a non-stationary fBm process with an exponent of $H-1$.

The EEG signals of the DEAP dataset provide a very low Hurst Exponent value that approaches $0$, while their profiles and separate bands provide an increased DFA-estimated exponent, still though below $0.2$ at most cases. The results however alter when we examine the profiles of the filtered bands, particularly alpha and beta, in which the exponent estimation shows a steady increase. These values confirm the evidence from the conducted ADF Test that EEG signals are negatively correlated and their fluctuations are smaller in larger time windows, which is the typical behavior of fGn processes having Hurst exponents below $0.5$.

\section{Feature Extraction}

\begin{figure}[]
    \centering
    \includegraphics[scale=0.63]{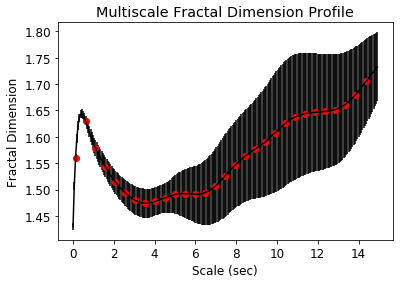}
    \vspace{-0.3cm}
    \caption{Sample MFD profile of a signal along with the mean and standard deviation features extracted from its 7 subsignals of 15 sec.}
    \vspace{-0.5cm}
\end{figure}

Each 60-sec EEG segment is partitioned to its main frequency bands through bandpass filtering with a 10th order Butterworth filter. We include alpha (8-13 Hz), beta (14-29 Hz), and gamma (30-45 Hz) rhythms, as well as raw signals in our analysis, since those have been acknowledged \cite{7104132} as the most emotion-sensitive. We select 12 left (Fp1, AF3, F7, F3, FC5, FC1, T7, C3, CP5, CP1, P3, P7) and 12 right (Fp2, AF4, F4, F8, FC2, FC6, C4, T8, CP2, CP6, P4, P8) channels that have shown competitive performance particularly when their asymmetrical relation is examined \cite{7104132}.

\vspace{-0.1cm}
\subsection{Baseline Features}

A set of widely used baseline features is extracted for comparative reasons and to assess the combined efficiency of the proposed feature set. These features are the Power Spectral Density (PSD) and the Higuchi Fractal Dimension (HFD) \cite{HIGUCHI1988277}. PSD is computed across the entire signals through the Welch method, resulting in 64 features per signal. The Higuchi Fractal Dimension is extracted using PyEEG \cite{pyeeg} and produces a scalar feature. To derive a feature vector here, we first split each signal into windows of 15 seconds (1920 samples) with 50\% overlap and then the HFD is determined for each one of the 7 windows, resulting in a 7D vector.

\begin{table*}[h]
\centering
\caption{Subject Dependent Task Accuracy in the form: Valence | Arousal}
\vspace{-0.15cm}
\begin{adjustbox}{width=0.87\textwidth,center}
\begin{tabular}{l||c|c|c|c|c|cc}
Features & Channels & Raw Signal & Alpha Band & Beta Band & Gamma Band & Combined \\ \hline \hline
\textbf{PSD} && \textbf{0.642} | 0.652   
    & 0.598 | 0.645    
    & 0.629 | 0.639   
    & \textbf{0.635} | 0.620 
    & 0.631 | 0.648 \\
\textbf{HFD} & Front & 0.615 | 0.638
    & 0.605 | 0.655    
    & 0.591 | 0.643   
    & 0.601 | 0.634 
    & \textbf{0.638} | 0.645 \\
\textbf{MFD} & Left & 0.620 | 0.661 
    & \textbf{0.626} | \textbf{0.669}    
    & 0.591 | \textbf{0.653}   
    & 0.594 | 0.636 
    & 0.612 | \textbf{0.661} \\
\textbf{MFDFA} && 0.577 | \textbf{0.662}     
      & 0.571 | 0.643    
      & 0.577 | 0.649   
      & 0.592 | \textbf{0.651} 
      & 0.586 | 0.658 \\ \hline
\textbf{PSD} && 0.627 | 0.644     
    & 0.616 | 0.645    
    & \textbf{0.637} | 0.641   
    & 0.623 | 0.627 
    & 0.623 | 0.646 \\
\textbf{HFD} & Front & 0.606 | 0.644     
    & 0.604 | 0.655    
    & 0.595 | 0.633   
    & 0.572 | 0.627 
    & 0.623 | 0.644 \\
\textbf{MFD} & Right & 0.607 | 0.655     
    & 0.605 | 0.652    
    & 0.566 | 0.652   
    & 0.602 | 0.641 
    & 0.597 | 0.657 \\
\textbf{MFDFA} && 0.587 | 0.655    
      & 0.573 | 0.641    
      & 0.603 | 0.650   
      & 0.573 | 0.620 
      & 0.586 | 0.652  \\
\end{tabular}
\end{adjustbox}
\vspace{-0.25cm}
\end{table*}

\begin{table*}[h]
\centering
\caption{Subject Independent Task Accuracy in the form: Valence | Arousal}
\vspace{-0.15cm}
\begin{adjustbox}{width=0.87\textwidth,center}
\begin{tabular}{l||c|c|c|c|c|cc}
Features & Channels & Raw Signal & Alpha Band & Beta Band & Gamma Band & Combined \\ \hline \hline
\textbf{PSD} && 0.554 | 0.569     
    & 0.547 | 0.564    
    & 0.549 | 0.562   
    & 0.553 | 0.570 
    & 0.546 | 0.564 \\
\textbf{HFD} & Front & 0.541 | 0.601     
    & 0.552 | 0.588    
    & 0.541 | 0.616   
    & 0.545 | 0.584 
    & \textbf{0.585} | \textbf{0.621} \\
\textbf{MFD} & Left & 0.553 | 0.606     
    & \textbf{0.566} | \textbf{0.631}    
    & 0.545 | \textbf{0.618}   
    & \textbf{0.554} | 0.580 
    & 0.559 | 0.615 \\
\textbf{MFDFA} && \textbf{0.569} | \textbf{0.630}
      & 0.546 | 0.600    
      & 0.545 | 0.598   
      & 0.532 | 0.545 
      & 0.553 | 0.608 \\  \hline
\textbf{PSD} && 0.553 | 0.580    
    & 0.557 | 0.560    
    & \textbf{0.558} | 0.573   
    & 0.552 | 0.579 
    & 0.555 | 0.575 \\
\textbf{HFD} & Front & 0.525 | 0.573     
    & 0.566 | 0.582    
    & 0.544 | 0.595   
    & 0.549 | 0.567 
    & 0.571 | 0.605 \\
\textbf{MFD} & Right & 0.552 | 0.601    
    & 0.556 | 0.605    
    & 0.547 | 0.587   
    & 0.545 | \textbf{0.588} 
    & 0.560 | 0.607 \\
\textbf{MFDFA} && 0.555 | 0.619    
      & 0.552 | 0.580    
      & 0.549 | 0.591   
      & 0.539 | 0.584 
      & 0.544 | 0.599  \\
\end{tabular}
\end{adjustbox}
\vspace{-0.5cm}
\end{table*}

\subsection{MFD Features}

Since MFD is mainly used for short-time analysis \cite{6365763}, we again split each signal into 15 sec. windows with 50\% overlap. The proposed feature set includes 30 linearly sampled features extracted out of each window's MFD. The respective features of each window are then summarized using 3 statistical metrics: mean, median and standard deviation. In this way, we get a final 90D feature vector incorporating the signal's temporal variance. The signals are analyzed at discrete scales of $s=1,...,274$ samples, thus the maximum scale is at $s = 1/7$ of the signals’ length. The fractogram of a sample signal along with the variance of its 7 windows are shown in Fig.~2. The EEG fractograms reveal a highly fragmented structure and a high fractal dimension $D>1.5$. This finding is consistent with the low Hurst Exponent we got from monofractal DFA.

\begin{figure}[t!]
    \centering
    \includegraphics[scale=0.6]{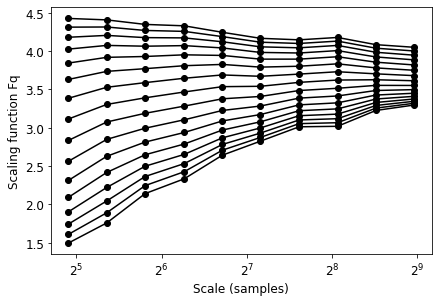}
    \vspace{-0.4cm}
    \caption{MFDFA on an EEG: depicting 16 linear-like graphs for
    $F_q(s)$ }
    \vspace{-0.5cm}
\end{figure}

\subsection{MFDFA Features}

We additionally acquire 30 features from processing the last half of each EEG waveform through the computationally expensive MFDFA. We select 10 scales ranging from 30 to 500 samples and $16$ q-moment values ranging from $-5$ to $5$. The resulting representation is a set of 16 linear-like graphs of 10 values, as shown in Fig~3. 16 Hurst Exponent values are determined through linear regression, one for each moment. The mass exponent $t$ is then derived through $t(q) = qH(q)-1$. A monofractal signal with constant $H$ would produce a linear graph, the EEG instead produces a curve that we utilize to produce the signal's \textit{multifractal spectrum} $D$:
$$ D(q) = q'h(q)-t(q'), \quad h(q_n) = \frac{t(q_n)-t(q_{n-1})}{q_n - q_{n-1}},$$
where $n=1,...,15$, $q'$ excludes the largest moment value, and $h(q)$ is the singularity exponent. The resulting curve, determined by 15 $h(q)$ and 15 $D(q)$ values, represents the MFDFA feature set.

\section{Experimental Evaluation}
\label{sec:page}

\textbf{Experimental Protocol:} We evaluate the features extracted from the multifractal analysis on the emotion recognition task. The experimental protocol can be divided into two categories: \textit{Subject Dependent}, in which a classifier is trained and tested on trials of a single participant, with the final score being the average per-subject score, and \textit{Subject Independent}, where a classifier is trained on several participants and tested against unseen trials. We shall mention that lower scores are typically reported \cite{eeg_general} for the latter, since EEG is highly subject-sensitive, thus we anticipate such behavior also in our models.

In this work, we make use of a single classifier unifying features from all available EEG channels. The model consists of a Standard Scaler, that standardizes training features by removing their mean and scaling them to unit variance, and a Support Vector Machine (SVM) with an RBF kernel. Experiments consider single labels, i.e., valence or arousal, in binary format by setting the threshold for binarization in the median score 5. We perform 5-fold cross validation on stratified splits of the available data: approximately 56.5\% of all samples are of high valence and 59\% of high arousal annotations.

\textbf{Comparison to Baselines:} The classification results for all features at the 2 distinct settings are summarized in Table~1 and 2. We notice the accuracy difference between subject dependent and independent tasks, supporting the claim that brain responses inherit mainly subjective characteristics. The EEG PSD is shown to be efficient in the subject-dependent setting, where the raw signal modality achieves 64.2\% in valence and 65.2\% in arousal. Interestingly, these scores significantly drop in the subject-independent setting, where the PSD emerges as the least efficient feature set, achieving only chance-level scores in arousal, 6\% below the top recorded accuracy of MFD. We can therefore assume that the within-subject variability is concentrated more on separate spectral characteristics of each participant and therefore, fractal analysis is more robust across subjects.

Multifractal methods show indeed strong performance in both experiments, surpassing chance levels and the baseline features in most cases. In contrast to spectral features that are sensitive to valence, these features prove efficient mainly in recognizing the arousal state, in which they achieve around 5\% higher scores. At the subject-dependent experiment particularly, MFD of the alpha band and MFDFA at the raw signal yield 66.9\% and 66.2\% respectively, whereas their highest subject-independent accuracy hits 63\%. Our results are in accordance with those reported in \cite{comp1} for PSD and HFD, while the top scores obtained by MFD and MFDFA surpass most of the ones reported there. On the other hand, at subject-independent classification, multifractal features perform comparably to those discussed at \cite{Cimtay_2020} for DEAP, although we recognize the additional difficulty of eliminating all of the trials of a tested participant from training.

\begin{table}[!t]
\vspace{-0.1cm}
\centering

\caption{MFD-HFD Arousal Accuracy}
\vspace{-0.15cm}
\begin{adjustbox}{width=0.5\textwidth,center}
\begin{tabular}{l||c|c|c|c|c|cc}
Feat/s & Exp & Raw & Alpha & Beta & Gamma & Comb \\ \hline \hline
\textbf{Left} & Subject & \textbf{0.663}
    & \textbf{0.670}    
    & \textbf{0.657}   
    & \textbf{0.637} 
    & 0.656 \\
\textbf{Right} & Dep. & 0.654     
    & \textbf{0.662}    
    & 0.618   
    & 0.640 
    & 0.655 \\ \hline
\textbf{Left} & Subject & \textbf{0.613}    
    & \textbf{0.641}    
    & 0.612   
    & 0.580 
    & 0.614 \\
\textbf{Right} & Indep. & \textbf{0.604}    
    & \textbf{0.610}    
    & 0.591   
    & 0.582
    & \textbf{0.615} \\
\end{tabular}
\end{adjustbox}
\vspace{-0.5cm}
\end{table}

\textbf{Aggregated features:} Multifractal features, and MFD in particular, clearly outperform the Higuchi baseline in arousal and perform comparably in valence, indicating that the multiscale variability of the EEG can capture latent emotional information. Furthermore, the two kinds of fractal dimensions provide even better scores in arousal when combined. As shown in Table~3, in both subject dependent and independent settings we record higher accuracy than the one we obtained from the individual features in the ablation study, mainly when testing raw signals or the alpha band. The differences are  significant in the subject independent setting, whereas even the top scores obtained previously are improved. The model can now predict arousal at 67\% and 64\% at subject dependent and independent experiments respectively.

Other than that, it seems that the selection of a single feature type could be adequate for affective state recognition, since we do not observe substantial improvement for the ``Combined'' features, in which we measure the aggregated performance of the three bands and the raw signal. We then compare the two asymmetrical sets of channels, where the left hemisphere is more efficient in terms of multifractal analysis, both in subject dependent and independent setting, performing 2\% better on average. In order to assess the asymmetrical performance, we also experimented on aggregated trials, the results however did not meet the scores obtained individually. It is evident though that higher accuracy is acquired when we consider the difference or quotient of those features, instead of just concatenating them. Finally, except for MFD and HFD, aggregated sets between the mentioned feature types do not provide a statistically significant improvement in recognition. We shall only mention that some valence scores including PSD features surpass some of those we report in the ablation study.

\section{Conclusion}

In this paper we analyzed the multiscale fractal structure of EEG and proposed a feature extraction method utilizing two multifractal algorithms for emotion recognition. The proposed features perform strongly against the baselines, particularly in the challenging subject-independent setting and in arousal recognition, indicating that arousal is correlated with the fragmented structure of the EEG. Further improvements are achieved when the fractal dimension features are aggregated, while the efficiency of the alpha frequency band is underlined in all experiments. Our analysis showed that multifractality and the anti-correlation properties should be considered when processing EEG signals. Further work on EEG emotion analysis should consider feature extraction algorithms for determining asymmetrical multifractal properties, whereas an interesting direction would be the examination of the correlation between brain signals and their stimuli.

\nocite{*}
\bibliographystyle{IEEEbib}
\bibliography{refs}
\end{document}